\documentclass[12pt]{iopart}
\usepackage{iopams}
\usepackage{graphicx}
\usepackage{color}

\begin{document}

\newcommand{\be}{\begin{equation}}
\newcommand{\ee}{\end{equation}}
\newcommand{\bea}{\begin{eqnarray}}
\newcommand{\eea}{\end{eqnarray}}
\newcommand{\nn}{\nonumber}
\newcommand{\de}{\partial}

\def\a{\alpha}
\def\b{\beta}
\def\d{\delta}        \def\D{\Delta}
\def\e{\epsilon}
\def\eps{\varepsilon}
\def\f{\phi}          \def\F{\Phi}
\def\vf{\varphi}
\def\g{\gamma}        \def\G{\Gamma}
\def\h{\eta}
\def\i{\iota}
\def\j{\psi}          \def\J{\Psi}
\def\k{\kappa}
\def\lam{\lambda}   \def\L{\Lambda}
\def\m{\mu}
\def\n{\nu}
\def\o{\omega}   \def\O{\Omega}
\def\p{\pi}      \def\P{\Pi}
\def\q{\theta}   \def\Q{\Theta}
\def\r{\rho}
\def\s{\sigma}   \def\S{\Sigma}
\def\t{\tau}
\def\u{\upsilon}  \def\U{\Upsilon}
\def\x{\xi}      \def\X{\Xi}
\def\z{\zeta}

\title[Resonant VE electron--N$_2$ and --NO cross sections]{Theoretical vibrational--excitation cross sections and rate coefficients for electron--impact resonant collisions involving rovibrationally excited N$_2$ and NO molecules}

\author{
V. Laporta$^{1,2,a}$\footnote[0]{$^a$ v.laporta@ucl.ac.uk},
R. Celiberto$^{2,3,b}$\footnote[0]{$^b$ r.celiberto@poliba.it} and
J. M. Wadehra$^{4,c}$\footnote[0]{$^c$ wadehra@wayne.edu}
}

\address{$^1$ Department of Physics and Astronomy, University College London, London WC1E 6BT, UK}
\address{$^2$ Institute of Inorganic Methodologies and Plasmas, CNR, 70125 Bari, Italy}
\address{$^3$ Department of Water Engineering and Chemistry, Polytechnic of Bari, 70125 Bari, Italy}
\address{$^4$ Physics Department, Wayne State University, Detroit, MI, 48202 USA}

\begin{abstract}
Electron--impact $v_i\rightarrow v_f$ vibrational excitations cross sections, involving rovibrationally excited N$_2(v_i,J)$ and NO$(v_i,J)$ molecules (fixed $J$), are calculated for collisions occurring through the nitrogen resonant electronic state N$_2^-(\textrm{X }^2\Pi_g)$, and the three resonant states of nitric oxide NO$^-(^3\Sigma^-, ^1\Delta, ^1\Sigma^+$). Complete sets of cross sections have been obtained for all possible transitions involving 68 vibrational levels of N$_2(\textrm{X }^1\Sigma_g^+)$ and 55 levels of NO$(\textrm{X }^2\Pi)$, for the incident electron energy  between 0.1 and 10 eV. In order to study the rotational motion in the resonant processes, cross sections have been also computed for rotationally elastic transitions characterized by the rotational quantum number $J$ running from 0 through 150. The calculations are performed within the framework of the \textit{local complex potential} model, by using potentials energies and widths  optimized in order to reproduce the experimental cross sections available in literature. Rate coefficients are calculated for all the $(v_i,J)\rightarrow (v_f,J)$ transitions by assuming a Maxwellian electron energy distribution function in the temperature range from 0.1 eV to 100 eV.

All the produced numerical data can be accessed at
\\
\textcolor{blue}{http://users.ba.cnr.it/imip/cscpal38/phys4entry/database.html}.
\end{abstract}

\maketitle

\section{Introduction\label{sec:intro}}
Vibrational transitions, induced by electron impact in diatomic molecules, play a key role in affecting the chemistry and the physics of many molecular plasma~\cite{09Cap}. The most efficient low-energy process, leading to molecular vibrational excitation/de-excitation, occurs through a resonant mechanism, according to which the incident electron is momentarily captured by a neutral molecule, in general vibrationally excited, with the subsequent formation of a resonant negative molecular ion. Within a given time interval, the electron can acquire all, or part, of its original kinetic energy and may escape from the molecule giving rise, correspondingly, to elastic or inelastic vibrational excitations~\cite{12Lapo}.

We will focus our discussion on electron-impact vibrational-excitation collisions involving N$_2$ and NO molecules. The interest in these species relies on their relevance, as plasma components, in several fields of scientific and technological importance: from planetary atmosphere, including environmental and spacecraft re-entry problems, to nuclear fusion and industrial plasmas~\cite{00Cap, 10Kall}.

Resonant vibrational excitations (RVE), for these two molecules, occur \emph{via} to the following mechanisms:
\begin{equation}
\mathrm{N}_2(\textrm{X }^1\Sigma_g^+, v_i,J)+e\rightarrow \mathrm{N}_2^-(\textrm{X }^2\Pi_g)\rightarrow \mathrm{N}_2(\textrm{X }^1\Sigma_g^+,v_f,J)+e\,, \label{RVE N2}
\end{equation}
\begin{equation}
\mathrm{NO}(\textrm{X }^2\Pi, v_i,J)+e\rightarrow \mathrm{NO}^-(^3\Sigma^-, ^1\Delta, ^1\Sigma^+)\rightarrow \mathrm{NO}(\textrm{X }^2\Pi, v_f,J)+e\,. \label{RVE NO}
\end{equation}
In both cases, the two molecules are initially in their respective ground electronic state and in one of the associated vibrational levels, labeled by the quantum number $v_i$, ranging from 0 through 67 for N$_2$ and 54 for NO and a fixed rotational quantum number $J$. The capture of the incident electron leads to the formation of a resonant negative ion which, after \textit{autoionization}, decays back to some level  $v_f$ of the neutral molecule. NO$^-$ contributes to process (\ref{RVE NO}) with the three resonant electronic states $^3\Sigma^-, ^1\Delta, ^1\Sigma^+$, while process (\ref{RVE N2}) proceeds through the resonant $\textrm{X }^2\Pi_g$ state, responsible for the well-known 2.3 eV resonance peak in electron-N$_2$ collision experiments~\cite{62Schulz}.

Experimental measurements~\cite{62Schulz, 64Schulz, 71Come, 84Wong, 85Allan, 05Allan, 97Swee, 95Sun} and theoretical calculations~\cite{79Dube, 79Schn, 83Berm, 86Huo, 99Miha, 05Trev, 08Houf_I, 08Houf_II} of cross sections exist in literature for the above processes. However, despite the broad production of cross sections data, especially for nitrogen, all the previous works are restricted to excitations between the first few vibrational levels. This circumstance is particularly limiting for models of those plasma systems in which strong non-equilibrium conditions exist. In these plasma systems the vibrational distribution is no longer Boltzmann, so that the high vibrational levels can become over-populated. An appropriate example of non-equilibrium conditions is provided by the air plasma generated by the impact of a supersonic space vehicles during its re-entry phase in the earth atmosphere. During the air-vehicle interaction, the kinetic energy transferred to atmospheric components induces complex physical and chemical processes and may, in particular, activate the internal degrees of freedom of the molecular species whose densities can strongly deviate from Boltzmann distribution. A realistic model, aimed to predict the chemical and physical evolution of these plasmas, relies therefore on a detailed description of the state-to-state processes occurring at the molecular level, with particular consideration of the electron-molecule energy exchanges, whose role is of fundamental importance in determining the level population of the chemical species.

In this paper we present complete sets of cross section data for both the excitation processes (\ref{RVE N2}) and (\ref{RVE NO}), connecting all the vibrational levels of N$_2$ and NO molecules, and for an incident electron energies interval below 10 eV. We have also extended the calculations to rotationally elastic resonant excitations of both the molecules ($J\leq 150$ and $\Delta J = 0$).  The cross sections are evaluated by resorting to the \textit{local complex potential} (LCP) model as formulated in Ref.~\cite{08Celi}. The LCP approximation is derived from the more rigorous \textit{nonlocal} model~\cite{91Domcke}, which is presently the best available theory for the description of the resonant collisions. Comparison of the two models, however, reported for N$_2$ and NO in Refs.~\cite{83Berm} and~\cite{05Trev}, respectively, for some transitions, shows that the results of the two methods are in good agreement, and the observed differences, although significant for testing the models, are quite inconsequential for any practical purpose. This is a favorable circumstance from a computational point of view, as the LCP model is less demanding, in terms of computer times, compared to nonlocal counterpart, especially if we consider, for example, that the complete sets of vibrational data include 68$\times$68 and 55$\times$55 possible transition cross sections to be evaluated for each energy grid point. Beside the non negligible time--saving features, the LCP approximation also provides quite accurate results demonstrated, as shown later, by the good agreement of the calculated cross sections with the experimental measurements.

Cross sections provide the basic input data for a plasma kinetic model, from which the rate coefficients can be obtained once the electron energy distribution function, for the particular conditions of the plasma system, is known. In many situations, however, when the electrons can be considered thermalized at a given temperature $T_e$, the rate coefficients can be obtained by convoluting the cross sections with a Maxwellian energy distribution function, according to the equation:
\begin{equation}
	K_{v_i\rightarrow v_f}(T_e) = \sqrt{\frac{8}{m_e\pi}}\left(\frac{1}{kT_e}\right)^{3/2}\int^\infty _{\epsilon_{th}} dE E\ e^{-\frac{E}{kT_e}}\ \sigma_{v_i\rightarrow v_f}(E)\,, \label{rate coefficient}
\end{equation}
where $m_e$ is the electron mass, $k$ is the Boltzmann constant, $\epsilon_{th}=v_f-v_i$ is the threshold energy and $\sigma_{v_i\rightarrow v_f}(E)$ is the cross section for the $v_i \rightarrow v_f$ resonant vibrational excitation for electron energy $E$. Using the calculated cross sections for both processes (\ref{RVE N2}) and (\ref{RVE NO}), we have also calculated the corresponding rate coefficients, by Eq.~(\ref{rate coefficient}), for all the transitions and in the range of electron temperature from 0.1 eV up to 100 eV.

The organization of the paper is as follows. In the next Section we briefly describe the main equations of the LCP model and give some computational details. In Section~\ref{sec:compar} we compare the calculated cross sections with some available theoretical and experimental results of the literature, and in Section~\ref{sec:results} we illustrate few particular examples of the calculated cross sections and rate coefficients. In Section~\ref{sec:summary}, finally, some conclusive comments will be provided.

\section{Theoretical model and numerical details\label{sec:th_model}}

In this section only a brief account is given of the relevant equations of the LCP model. A comprehensive formulation of the resonant collisions can be found in~\cite{08Celi} and in references therein.

The cross section for a given rovibrational $(v_i,J_i)\rightarrow (v_f,J_f)$ transition, occurring through an electron-molecule resonant collision is given by~\cite{08Celi}:
\begin{equation}
	 \sigma_{if}(\epsilon_i) = \frac{g_r}{g}\ \frac{2\,\pi^3\,m_e^2}{\hslash^2}\ \frac{k_f}{k_i}\ \left|\langle \chi_f|V_{k_f}|\xi\rangle\right|^2\,, \label{xsec}
\end{equation}
where $g_r$ and $g$ represent, respectively, the product of the degeneracy factors and spin-multiplicity of the resonant and neutral electronic states. The equation includes a multiplicative factor of 1/2 in order to account for the spin-multiplicity of the unpolarized incident electron. The initial and final electron momentums are defined by $k_i^2=2m_e\epsilon_i/\hbar^2$ and $k_f^2=2m_e\epsilon_f/\hbar^2$, where $\epsilon_i$ and $\epsilon_f$ are the initial and the final kinetic energies of the free electron. $\chi_{v_f,J_f}(R)$, is the wave function of the final rovibrational state of the target molecule and $\xi(R)$ the nuclear wave function of the resonant state, both depending on the internuclear distance $R$. The \textit{exit amplitude} $V_{k_f}(R)$ is the value of the interaction matrix element, $V_{k}(R)$, coupling the discrete and continuum states of the colliding system~\cite{08Celi} for the exit channel $k=k_f$.

Once the coupling matrix $V_{k}(R)$ and the potential energy curve of the resonant state, $V^-(R)$, are known, the wave function $\xi(R)$ can  be obtained in the LCP model by solving the nuclear resonant equation:
\bea
\left[-\frac{\hbar^2}{2\mu }\frac{d^2}{dR^2}+\frac{J_i(J_i+1)\hbar^2}{2\mu R^2} + V^-(R) + \Delta (R)- \frac{i}{2}\Gamma(R) - E \right]\xi(R) =& & \nn
\\
\hspace{4cm} = -V_{k_i}(R)\ \chi_{v_i,J_i}(R) \,,&& \label{res wf}
\eea
where $\mu$ is the nuclei reduced mass and $E=\epsilon_i +E_{v_i,J_i}=\epsilon_f +E_{v_f,J_f}$ is the total energy. $E_{v,J_i}$ are the rovibrational eigenvalues of the target Schr\"{o}dinger equation:
\begin{equation}
	\left[-\frac{\hbar^2}{2\,\mu }\frac{d^2}{dR^2}+\frac{J_i(J_i+1)\hbar^2}{2\mu R^2} +V_0(R)\right]\chi_{v,J_i}(R) = E_{v,J}\chi_{v,J_i}(R)\,. \label{vibr. eq.}
\end{equation}
Here $V_0(R)$ is the electronic potential energy function of the neutral molecule. In Eq.~(\ref{res wf}), $\Gamma(R)$ and $\Delta(R)$ are respectively the \textit{resonance width} and the \textit{level shift}, and finally, $V_{k_i}(R)$ is the \textit{entry amplitude} describing the capture of the incident electron by the molecule.

\begin{figure}[ht!]
\begin{indented}
\item[]
\includegraphics[scale=.7,angle=0]{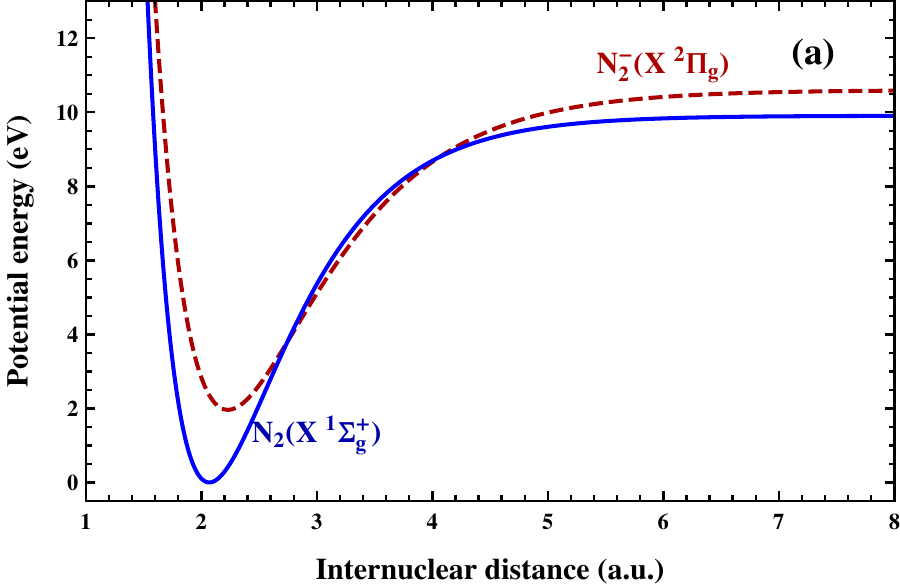}\hfill
\includegraphics[scale=.7,angle=0]{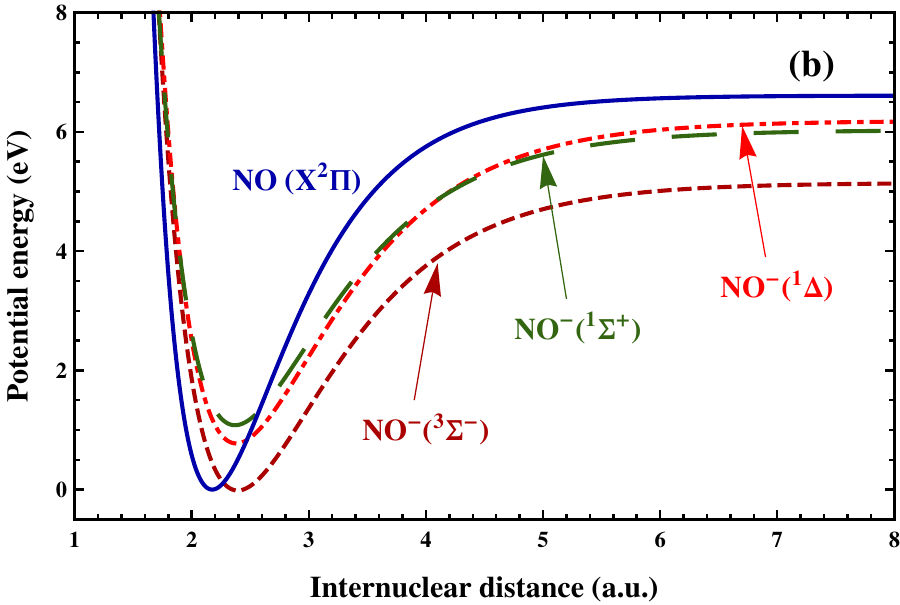}
\end{indented}
\caption{Morse potentials for: (a) N$_2$ (solid line) and N$_2^-$ (dashed line); (b) NO (solid curve) and the three NO$^-$ states: $^3\Sigma^-$ (short dashed line), $^1\Delta$ (dot-dashed line) and $^1\Sigma^+$ (long dashed line).}
\label{N2-NO potentials}
\end{figure}

The potential curves, $V^-(R)$ and $V_0(R)$, needed in Eq.~(\ref{res wf}) and~(\ref{vibr. eq.}) respectively, have been expressed as a Morse-like function:
\begin{equation}
	U(R) = D_e\left[1-e^{-\alpha(R-R_e)}\right]^2+W\,, \label{Morse f}
\end{equation}
where $D_e$ is the dissociation energy for the considered electronic state and $R_e$ is its equilibrium bond-length. $W$ is the zero-point energy and $\alpha$ is a constant.
\begin{table}[t!]
\begin{center}
\caption{\label{table: morsepar} Morse parameters for N$_2$, N$_2^-$, NO and NO$^-$ potential states (Eq.~(\ref{Morse f})) and reduced mass $\mu$ for the neutral species.}
\begin{tabular}{|c|c|c|c|c|c|c|}
\hline
& N$_2(\textrm{X }^1\Sigma^+_g)$ & N$_2^-(\textrm{X }^2\Pi_g)$ & NO$(\textrm{X }^2\Pi)$ & NO$^-(^3\Sigma^-)$ & NO$^-(^1\Delta)$ & NO$^-(^1\Sigma^+)$
\\
\hline
$D_e$ (eV)        & 9.906  & 8.638 & 6.61  & 5.161   & 5.411 & 4.95
\\
$R_e$ (a.u.)      & 2.069  & 2.230 & 2.175  & 2.393  & 2.38  & 2.37
\\
$\alpha$ (a.u.)   & 1.428  & 1.20  & 1.48  & 1.20    & 1.18  & 1.20
\\
$W$ (eV)          & 0      & 2.18  & 0     & -0.015   & 0.775 & 1.08
\\
$\mu$ (a.u.)      & 12853  &       & 13614 & &&
\\
\hline\hline
\end{tabular}
\end{center}
\end{table}
The Morse potential parameters for the nitrogen molecule were obtained, by a fitting procedure, from the Gilmore potential~\cite{65Gilm}. The dissociation energy, measured from the potential bottom, has been taken from Ref.~\cite{50Herz}, while $\alpha$ is set in order to reproduce the first few vibrational levels of Ref. \cite{90Liu}. For NO and NO$^-$ the Morse parameters were taken from the Refs.~\cite{68Spen, 77Teil, 88deVie}. All the constants entering in Eq.~(\ref{Morse f}) for the electronic states involved in the processes (\ref{RVE N2}) and (\ref{RVE NO}) are given in Table~\ref{table: morsepar}, along with the reduced mass, $\mu$, of N$_2$ and NO molecules. The potential curves, obtained in this manner, are shown in Fig.~\ref{N2-NO potentials} (a)-(b).

\begin{figure}
\begin{indented}
\item[]
\includegraphics[scale=.7,angle=0]{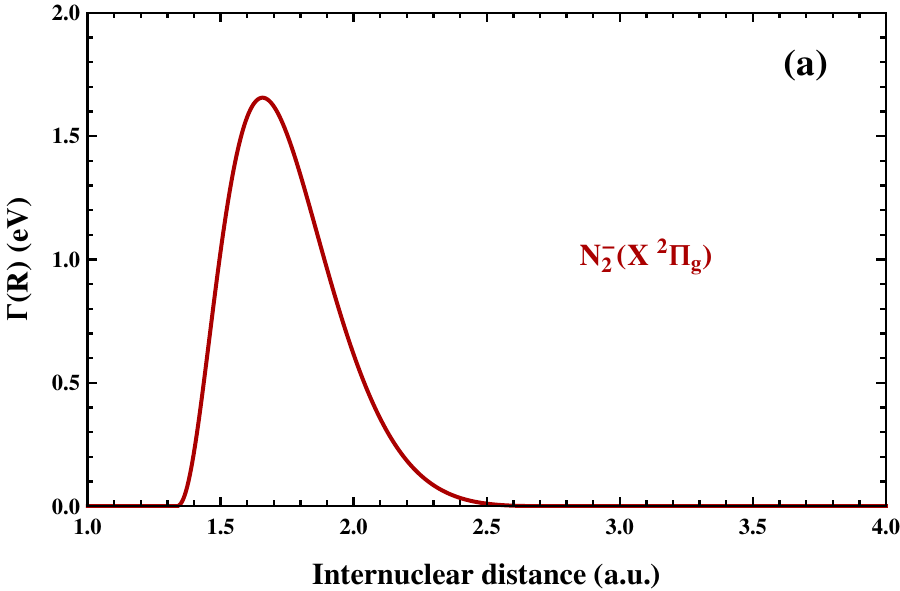}\hfill
\includegraphics[scale=.7,angle=0]{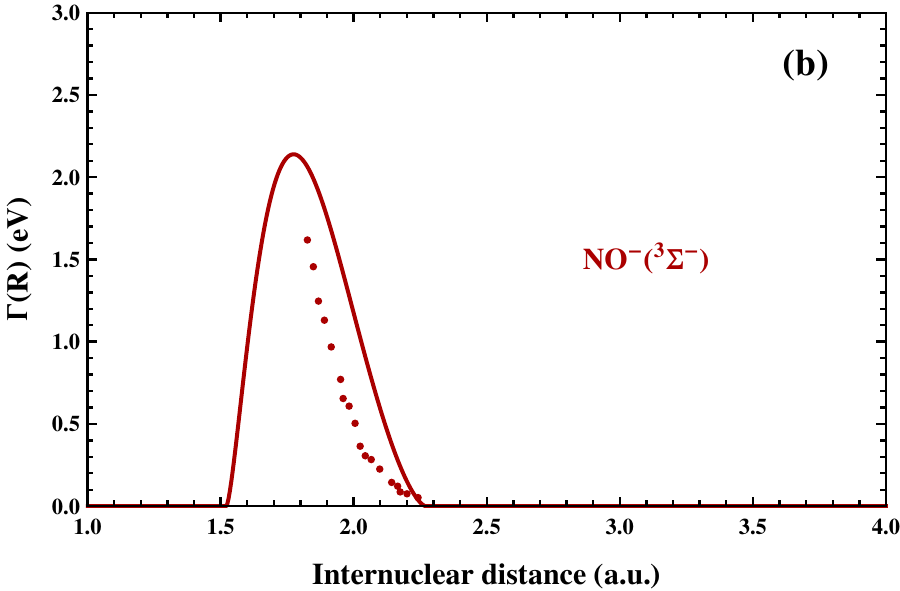}
\\
\includegraphics[scale=.7,angle=0]{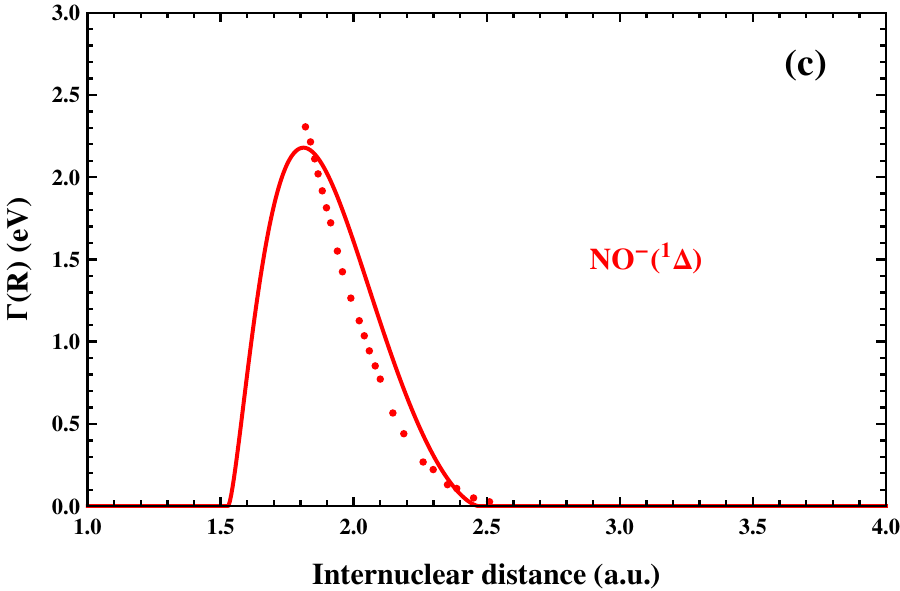}\hfill
\includegraphics[scale=.7,angle=0]{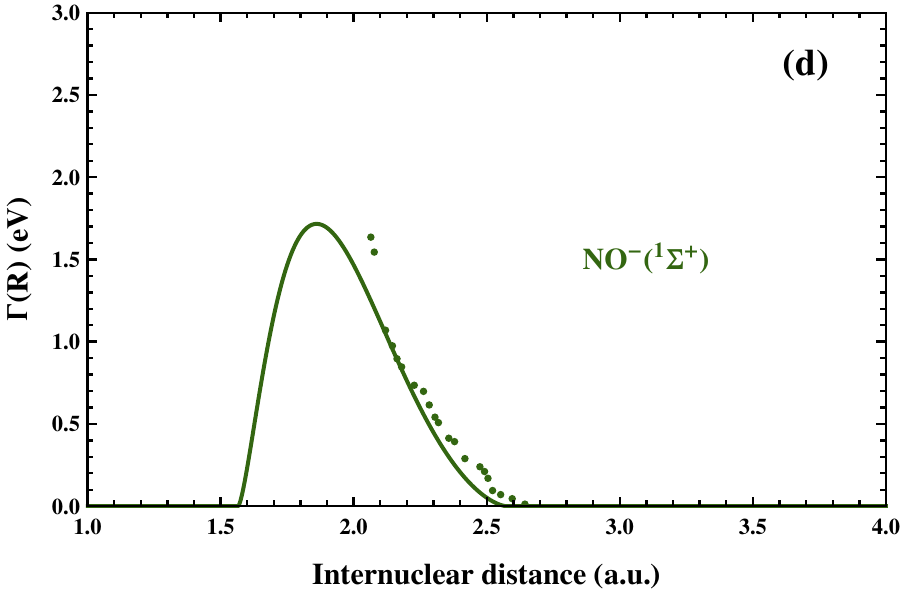}
\end{indented}
\caption{Resonance width obtained according to Eq.~(\ref{eq:gamma}) for N$_2^-$ and for the three states of NO$^-$ as indicated in the figures. The points refer to R-matrix calculation taken from Ref.~\cite{04Zhang}.}
\label{fig: gamma}
\end{figure}

The width function $\Gamma(R)$ in Eq.~(\ref{res wf}) has been expressed as~\cite{79Dube}:
\begin{equation}
 \Gamma(R) = \sum_i\ C_i\ \left[V^-(R)-V_0(R)\right]^{l_i+\frac12}\ H\left(V^--V_0\right)\,, \label{eq:gamma}
\end{equation}
where $H$ is the Heaviside step--function and $l_i$ is the angular momentum of the lowest contributing partial wave associated with the incident electron. The sum in Eq.~(\ref{eq:gamma}) runs over all the electronic states of the molecular negative ion involved in the process, which implies one term for N$_2^-$ and three for NO$^-$. For the corresponding value of the angular quantum number we have $l_1=2$ for nitrogen, due to the $d$-wave nature of the lowest partial wave of the incident electron~\cite{83Berm}, and $l_1=l_2=l_3=1$ for NO, corresponding to a $p$-wave nature for the three resonant states~\cite{04Zhang}. In order to reproduce the position and width of the peaks in the experimental cross section, the constants $C_i$ in Eq.~(\ref{eq:gamma}), for $\Gamma$, were considered as phenomenological external parameters of the model and empirically adjusted. Their values are given in Table~\ref{table: gamma_and_delta} for both the processes (\ref{RVE N2}) and (\ref{RVE NO}). The shift operator $\Delta(R)$, in Eq.~(\ref{res wf}), has been included in the resonant potential~\cite{08Celi}. The plots in Fig.~\ref{fig: gamma} show the resonance widths for N$_2^-$ and NO$^-$, obtained using Eq.~(\ref{eq:gamma}). In the case of NO$^-$ the calculated widths are compared with the R-matrix results, where the resonance position and width are calculated by fitting the eigenphases sum in fixed-nuclei approximation with a Breit-Wigner profile~\cite{04Zhang}. To validate the numerical results,we changed the coefficients in Table~\ref{table: gamma_and_delta} by $\pm10\%$ and noted that the effect on the cross sections due to this change is very negligible, in particular for N$_2$ case.
\begin{table}[t!]
\begin{center}
\caption{\label{table: gamma_and_delta} Coefficients $C_i$, in Eq.~(\ref{eq:gamma}), for N$_2^-$ and NO$^-$ resonant states.}
\begin{tabular}{|c|c|c|c|c|}
\hline
& N$_2^-(\textrm{X }^2\Pi_g)$ & NO$^-(^3\Sigma^-)$ & NO$^-(^1\Delta)$ & NO$^-(^1\Sigma^+)$
\\
\hline
$C_i$ (eV$^{0.5-l_i}$)    & 0.05        & 0.81        & 0.60        & 0.50
\\
\hline\hline
\end{tabular}
\end{center}
\end{table}

The  discrete-continuum coupling matrix element $V_k(R)$ of Eq.~(\ref{res wf}) is written for nitrogen as~\cite{08Celi}:
\begin{equation}
V_k^2(R) = \frac{1}{2\pi}\frac{\Gamma(R)}{k(R)}\,, \label{eq:VkR}
\end{equation}
where $k(R)$ is given by
\begin{equation}
k^2(R) = \frac{2\,m}{\hbar^2}\left[V^-(R)-V_0(R)\right]\,. \label{eq:k_electron}
\end{equation}
For the case of NO we have modulated $V_k(R)$, following Ref.~\cite{05Trev}, with the so-called penetration factor, an \textit{ad hoc} function introduced to force the correct dependence of the cross section on the energy near the excitation threshold, namely,
\begin{equation}
  f(k,R) = \left\{
\begin{array}{ll}
k/k(R) & \textrm{if~} k<k(R)
\\
1 & \textrm{otherwise}\,.
\end{array}\right. \label{penetration factor}
\end{equation}
The coupling matrix element for NO is thus written as:
\begin{equation}
V^2_k(R) = \,f^{2l+1}\,\frac{1}{2\pi}\frac{\Gamma(R)}{k(R)}\,.
\end{equation}
\ \

Eq.~(\ref{xsec}) can be related to differential cross section by the approximate expression:
\begin{equation}
	\sigma_{if}(\epsilon_i) = \frac{1}{g_{l_i}(\theta)}\frac{d\sigma_{if}(\Omega)}{d\Omega}\,, 	 \label{diff_xsec}
\end{equation}
where the angular factor $g_{l_i}$, depending on the scattering angle $\theta$, is explicitly defined in Ref.~\cite{08Celi}. In the following, for purposes of comparison, we will make use of this equation only to extract the integral cross section $\sigma_{if}(\epsilon_i)$ from the measured differential data.

In the present paper we have computed the cross sections for the transitions $(v_i, J_i)\rightarrow (v_f,J_f)$ involving all the possible vibrational levels of N$_2$ and NO molecules and for all the diagonal rotational transitions $J_i=J_f=J$ with $0\leq J\leq 150$.

\section{Theoretical and experimental comparisons\label{sec:compar}}
In order to ascertain the accuracy of LCP calculations we compare our resonant vibrational excitation (RVE) cross section results for the N$_2$ and NO molecules with the theoretical and experimental cross sections and rate coefficients presently available in literature.

\subsection{\rm{N}$_2$}
The cross sections for process (\ref{RVE N2}) have been calculated by Eq.~(\ref{xsec}) where the degeneracy factors for N$_2(\textrm{X }^1\Sigma_g^+)$ and N$_2^-(^2\Pi_g)$ have been set to $g=1$ and $g_r=4$ respectively.

\begin{figure}
\begin{indented}
\item[]
\includegraphics[scale=.7,angle=0]{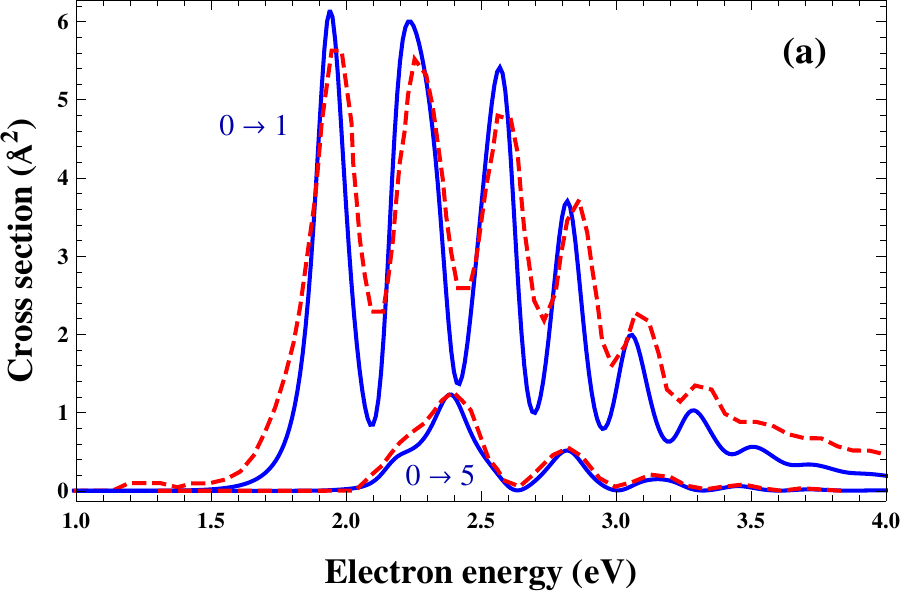}\hfill
\includegraphics[scale=.73,angle=0]{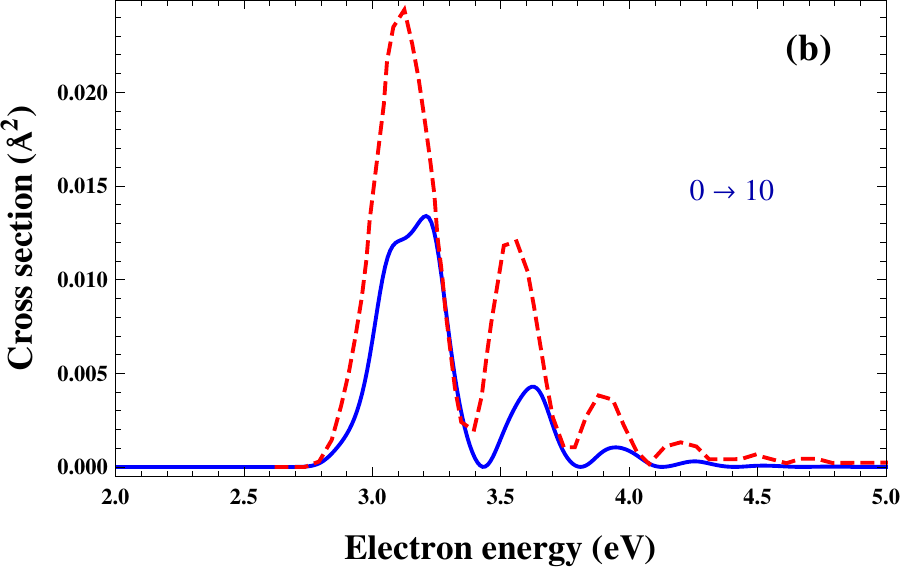}
\end{indented}
\caption{Electron-N$_2$ theoretical-experimental cross section comparison for the rotationless $v_i=0\rightarrow v_f=1,5,10$ transitions as indicated in the panels. Full lines: present calculations; dashed lines: experimental results of Allan~\cite{85Allan}.}
\label{fig: Allan}
\end{figure}

Fig.~\ref{fig: Allan} compares the cross sections for the rotationless $v_i=0\rightarrow v_f=1,5,10$ transitions with the experimental data of Allan~\cite{85Allan}. 
The comparison shows a very good agreement among the two sets of data for the first two transitions, while some discrepancy, around a factor of 2.8 at main peak, is observed in Fig.~\ref{fig: Allan} (b) for the $0\rightarrow 10$ excitation. In this last case, however, the absolute cross section values are relatively quite small, which can lead to significant uncertainties in experimental values.

\begin{figure}
\begin{indented}
\item[]
\includegraphics[scale=.7,angle=0]{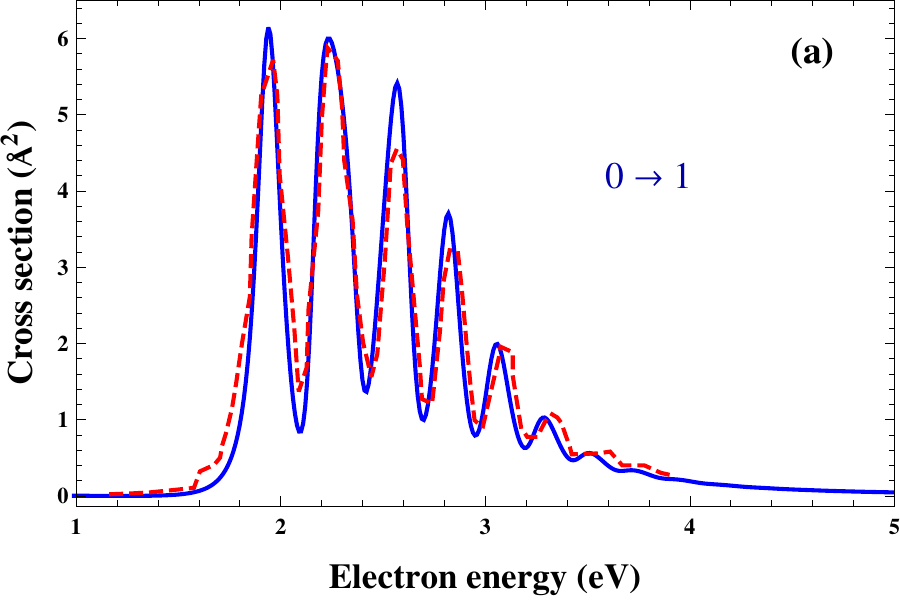}\hfill
\includegraphics[scale=.7,angle=0]{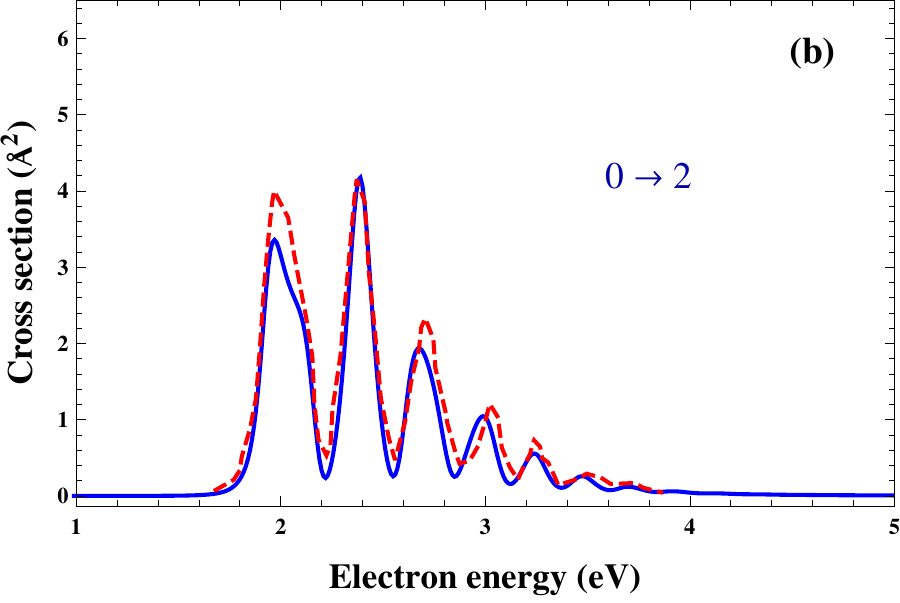}
\end{indented}
\caption{Electron-N$_2$ theoretical-experimental cross section comparison for the rotationless $v_i=0\rightarrow v_f=1,2$ transitions as indicated in the panels. Full lines: present calculations; dashed lines: unpublished experimental results of Wong as cited in Ref.~\cite{{79Dube}}.}
\label{fig: Wong}
\end{figure}

Fig.~\ref{fig: Wong} compares the theoretical RVE results with unpublished experimental measurements of Wong as reported in Ref.~\cite{79Dube}, for the rotationless vibrational transitions $0\rightarrow 1,2$. Our calculations are able to reproduce the experimental values of the number of peaks, their positions and peak values of cross sections quite accurately.

\begin{figure}
\begin{indented}
\item[]
\includegraphics[scale=.7,angle=0]{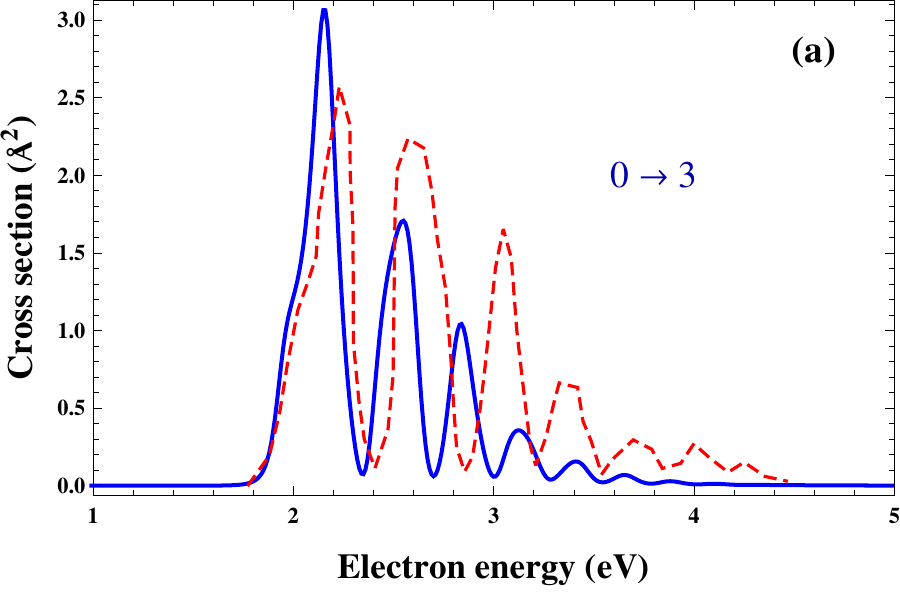}\hfill
\includegraphics[scale=.7,angle=0]{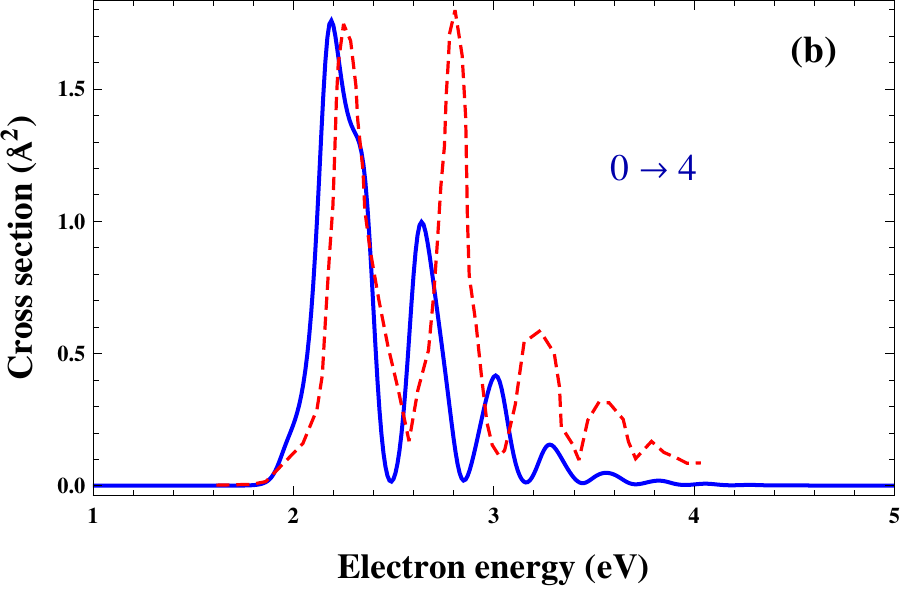}
\end{indented}
\caption{Electron-N$_2$ comparison of the present RVE calculations (full line) with the R-matrix results (dashed line)~\cite{79Schn} for the rotationless transitions shown in the panels.}
\label{fig: Rmatrix}
\end{figure}

Fig.~\ref{fig: Rmatrix} (a)-(b) shows a comparison of the present cross sections with those obtained with a non-adiabatic R-matrix approach~\cite{79Schn} for the rotationless vibrational transitions $0\rightarrow 3$ and $0\rightarrow 4$ respectively. The relative position of the peaks in the two calculations is shifted for both the transitions. This discrepancy can be ascribed to the different input quantities used in the two calculations. The RVE cross sections, defined earlier, are, in fact, quite sensitive to variations of the potential curves and widths~\cite{81Hazi}, and, as explained in Sec.~\ref{sec:th_model}, we have adjusted our input parameters to reproduce the experimental data of Allan~\cite{85Allan}.

\subsection{\rm{NO}}
For the RVE of NO molecule, we have calculated the cross sections separately for each of the three processes $^2\Pi\rightarrow ^1\Sigma^+$, $^2\Pi\rightarrow ^3\Sigma^-$ and $^2\Pi\rightarrow ^1\Delta$. The total cross section has been then obtained as a weighted sum of all the contributions, namely~\cite{05Trev},
\begin{equation}
	\sigma_{v_i \rightarrow v_f}(\epsilon_i) =\frac{1}{8}\left[\sigma^{^2\Pi\rightarrow ^1\Sigma^+}_{v_i \rightarrow v_f}(\epsilon_i) + 3 \sigma^{^2\Pi\rightarrow ^3\Sigma^-}_{v_i \rightarrow v_f}(\epsilon_i) + 2 \sigma^{^2\Pi\rightarrow ^1\Delta}_{v_i \rightarrow v_f}(\epsilon_i)\right]\,, \label{NO total xsec}
\end{equation}
where the coefficients are the statistical weights of each state~\cite{05Trev}. Interference terms are neglected.

\begin{figure}
\begin{indented}
\item[]
\includegraphics[scale=.7,angle=0]{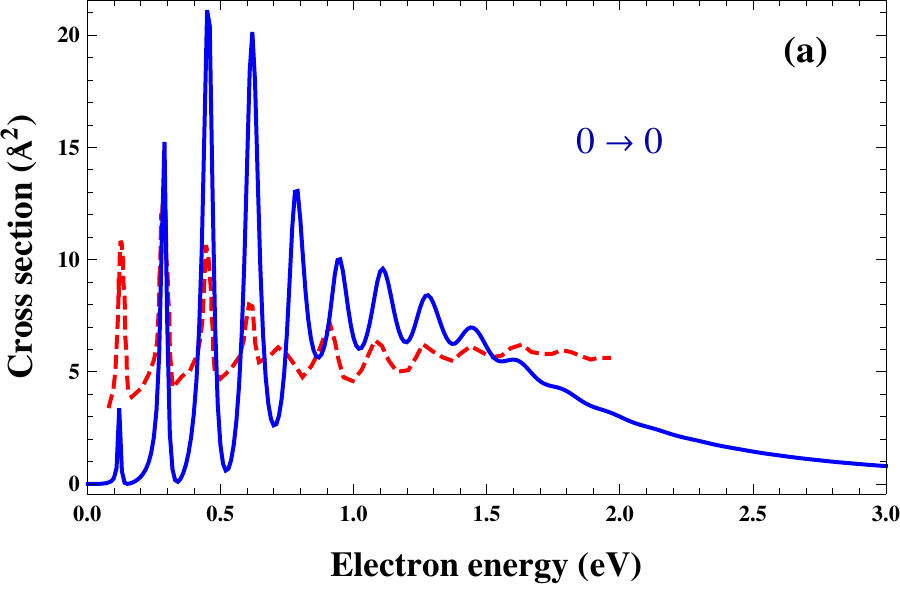}\hfill
\includegraphics[scale=.7,angle=0]{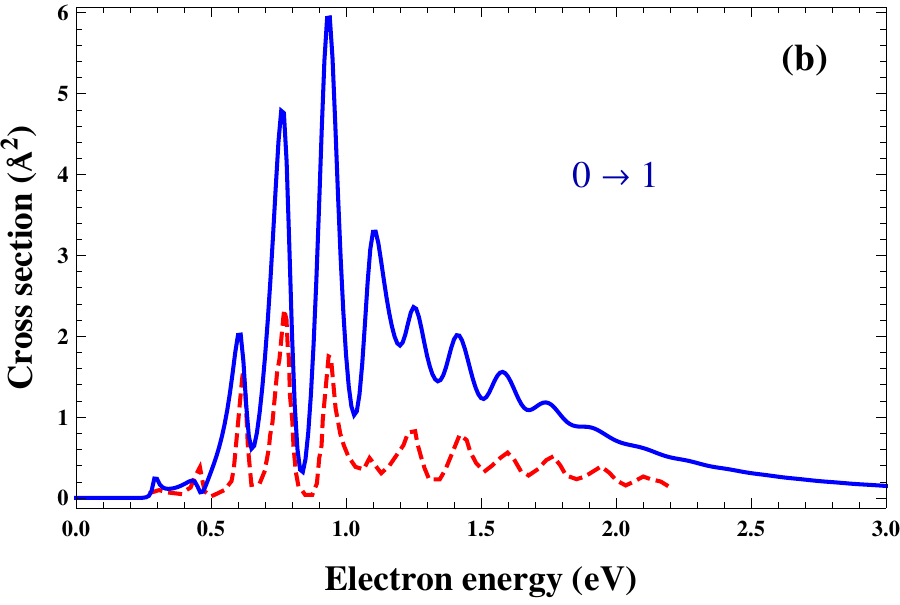}
\\
\includegraphics[scale=.7,angle=0]{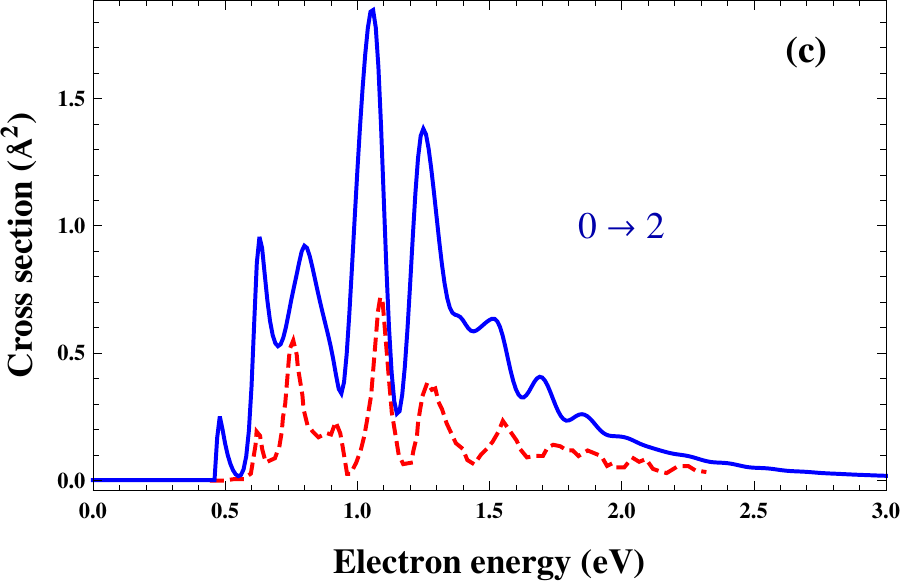}\hfill
\includegraphics[scale=.7,angle=0]{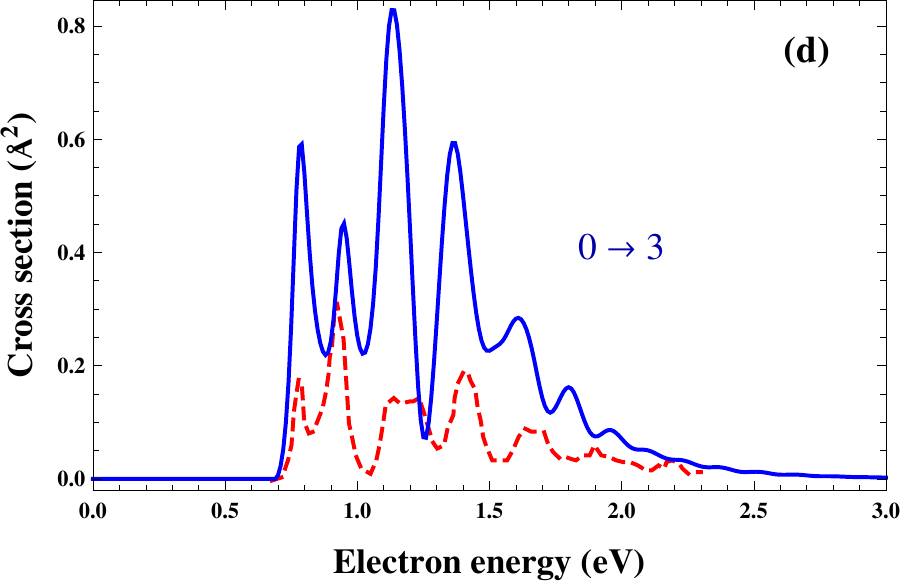}
\end{indented}
\caption{Electron-NO comparison of the present RVE calculations (full lines) with the measurements of ref.~\cite{05Allan} (dashed lines) multiplied by a factor of $4\pi$ (see text) for the vibrational transitions $0\rightarrow 0, 1, 2, 3$.}
\label{fig: NO-Allan}
\end{figure}

Fig.~\ref{fig: NO-Allan} shows a comparison of our RVE cross sections for the $0\rightarrow 0, 1, 2, 3$ vibrational transitions with the differential cross sections measured by Allan~\cite{05Allan} at an angle of 135$^\mathrm{o}$ and multiplied, according to Ref.~\cite{05Trev}, by a factor of $4\pi$ to get, approximately, the integrated cross sections. The agreement is quite satisfactory considering the fact that we used these experimental results to optimize the input parameters for the adiabatic potentials and resonance widths of the Table~\ref{table: morsepar} and~\ref{table: gamma_and_delta}. No nonresonant background contribution has been included in the calculations. In fact, in Fig.~\ref{fig: NO-Allan}~(a) the discrepancy between our calculated ($0\to 0$) elastic cross sections and the measurements of Ref.\cite{05Allan} is precisely due to the contribution of nonresonant background.

\begin{figure}
\begin{indented}
\item[]
\includegraphics[scale=.7,angle=0]{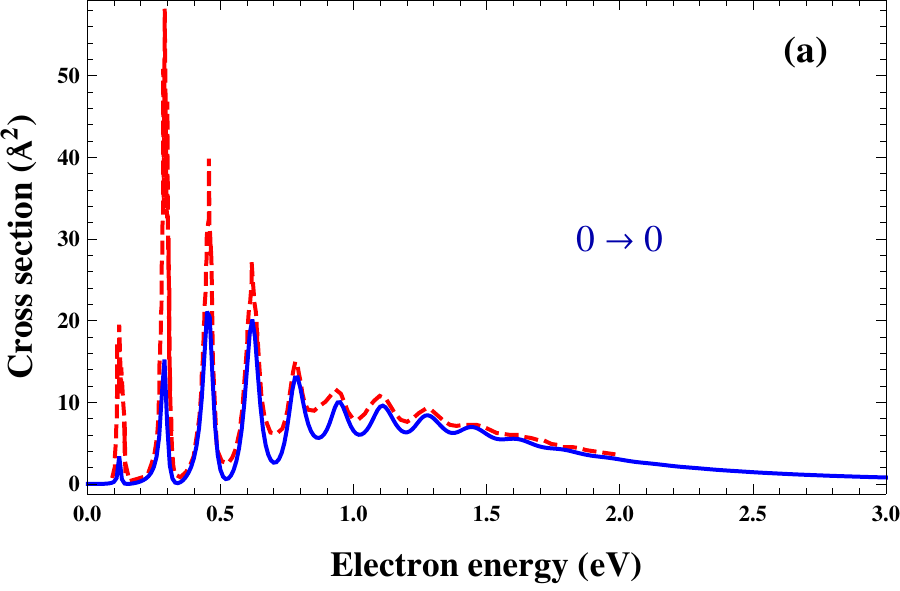}\hfill
\includegraphics[scale=.7,angle=0]{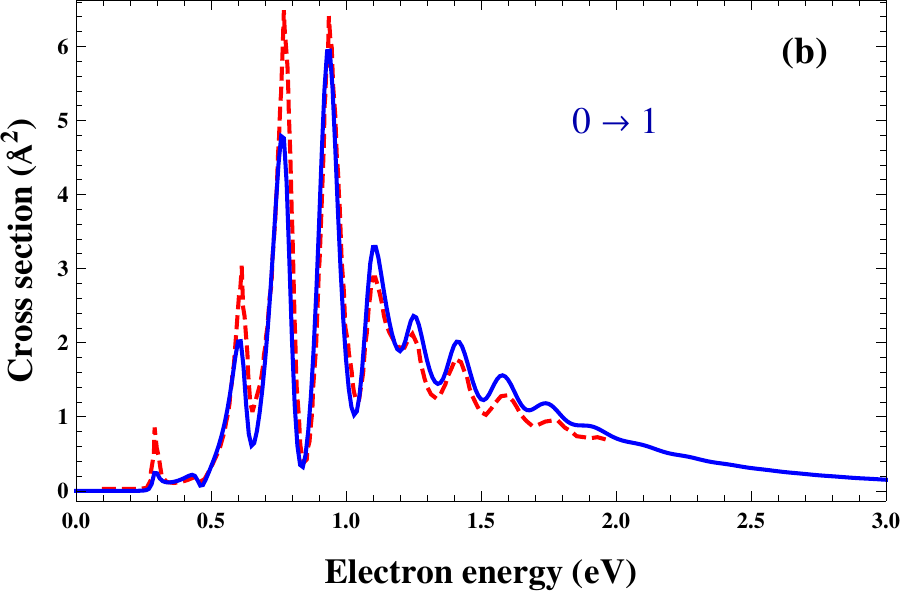}
\\
\includegraphics[scale=.7,angle=0]{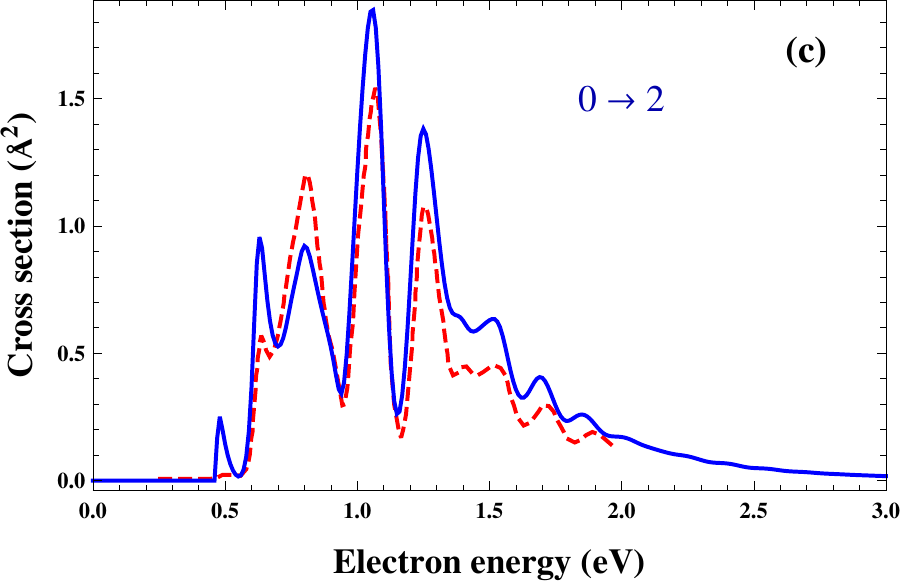}\hfill
\includegraphics[scale=.7,angle=0]{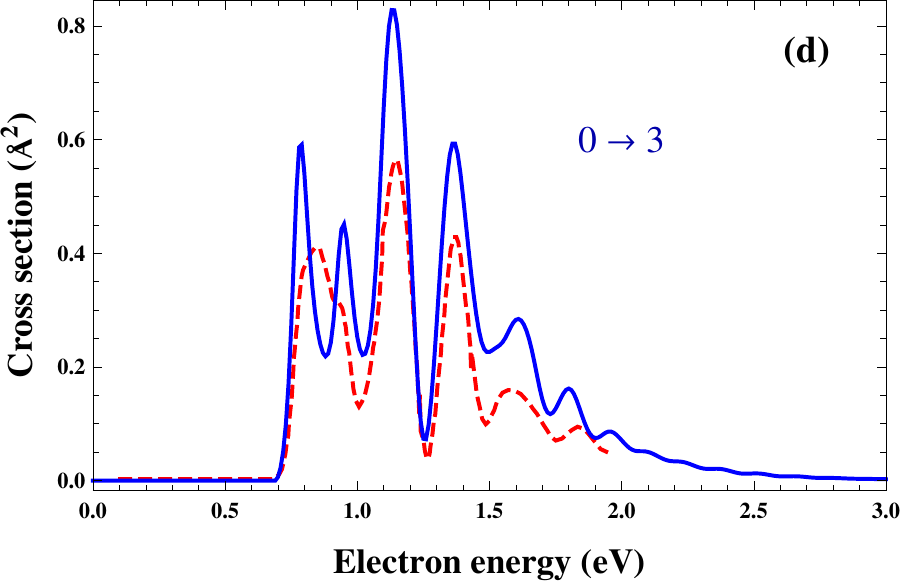}
\end{indented}
\caption{Electron-NO comparison of the present RVE calculations (full lines) with those of Trevisan \emph{et al.}~\cite{05Trev} (dashed lines) for the vibrational transitions $0\rightarrow 0, 1, 2, 3$.}
\label{fig: NO-Trevisan}
\end{figure}

Fig.~\ref{fig: NO-Trevisan} shows a comparison of our RVE cross sections with the non local calculations of Trevisan \emph{et al.}~\cite{05Trev} for the $0\rightarrow 0, 1, 2, 3$ vibrational transitions. The agreement here also is very satisfactory for all the transitions except for the elastic process, where the non local cross sections are significantly higher than the local results. A discrepancy by a factor of 6 is in fact observed at the second peak with respect to the present calculations which, however, have been calibrated using the experimental data of Allan.

\subsection{Rate coefficients}
\begin{figure}
\begin{indented}
\item[]
\includegraphics[scale=.7,angle=0]{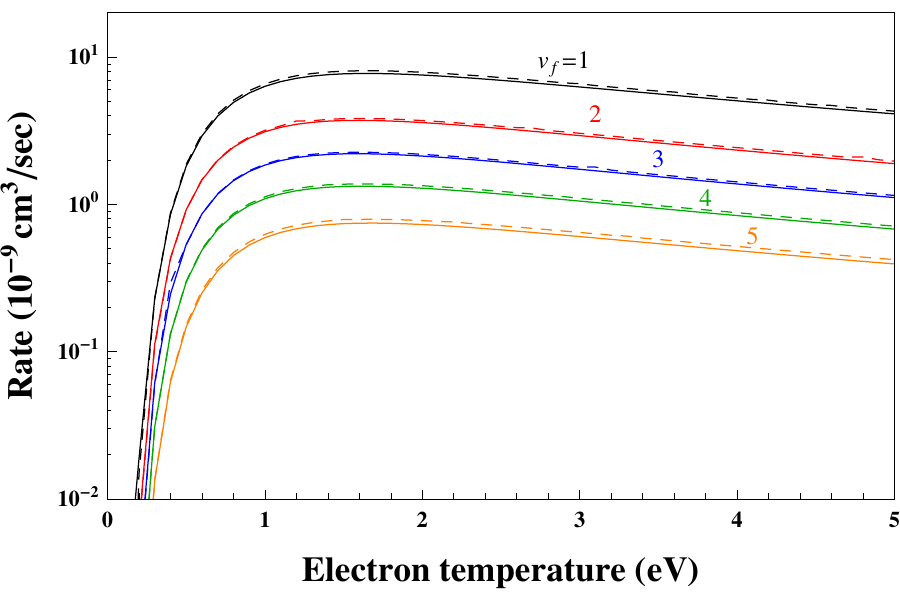}
\end{indented}
\caption{Comparison of the present electron-N$_2$ RVE rate coefficients (full lines) with those of Huo \emph{et al.}~\cite{86Huo} (dashed lines) as a function of the electron temperature. The curves refer to the transitions starting from $v_i=0$ and ending on the final levels $v_f=1,2,3,4,5$ as indicated in the figure.}\label{fig: NO-Huo}
\end{figure}

The rate coefficients for both the processes (\ref{RVE N2}) and (\ref{RVE NO}) were calculated using Eq.~(\ref{rate coefficient}) by assuming a Maxwellian electron energy distribution function. A comparison for N$_2+e$ system, with the well known rate coefficients of Huo~\cite{86Huo}, is shown in Fig.~\ref{fig: NO-Huo} for the $v_i=0\rightarrow v_f=1,2,3,4,5$ transitions up to an electronic temperature of $T_e=5$ eV. The representative curves almost overlap, so the two calculations give, in fact, almost the same results for the transitions shown in the figure.

To the best of our knowledge, no rate coefficient data for the electron collision processes involving the NO molecule, are available in literature.

\section{Results\label{sec:results}}
As mentioned earlier, we calculated the cross sections and rate coefficients as a function of the electron energy and electron temperature, respectively, for all possible electron-impact-induced rovibrational transitions occurring according to the processes (\ref{RVE N2}) and (\ref{RVE NO}). Since the database produced is prohibitively large to be completely shown in the present paper, in this Section we limit ourselves to the discussion of some particular cases. The whole database, however, will be provided in electronic form elsewhere (see Section~\ref{sec:summary}).

\begin{figure}
\begin{indented}
\item[]
\includegraphics[scale=.7,angle=0]{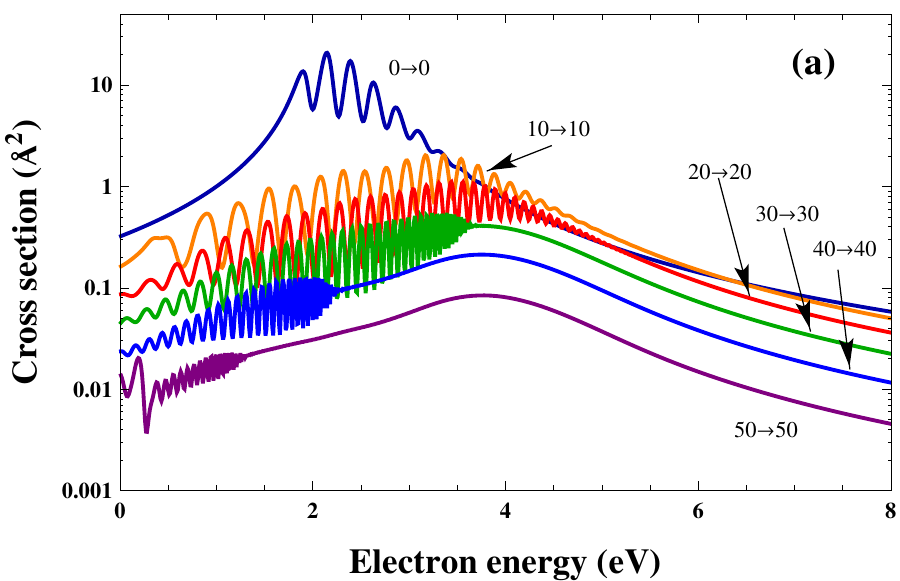}\hfill
\includegraphics[scale=.7,angle=0]{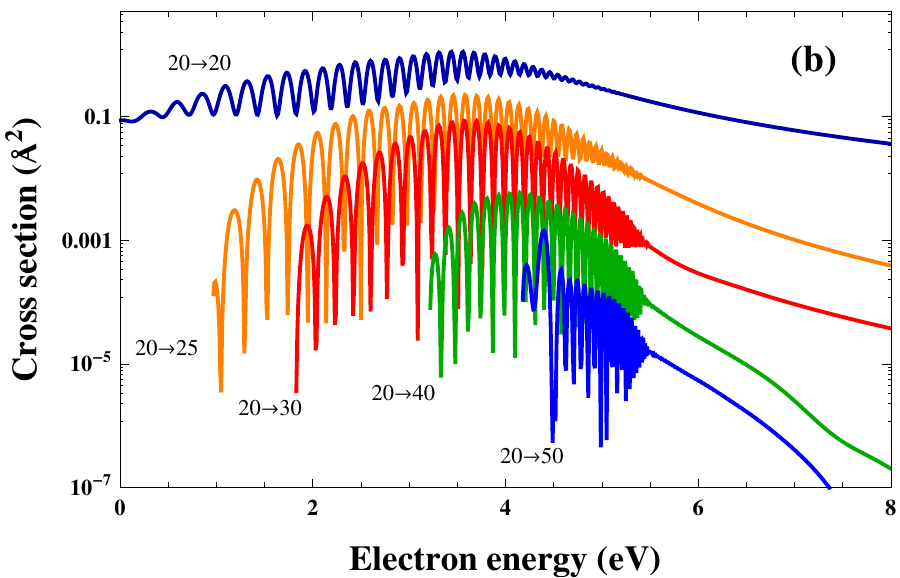}
\\
\includegraphics[scale=.7,angle=0]{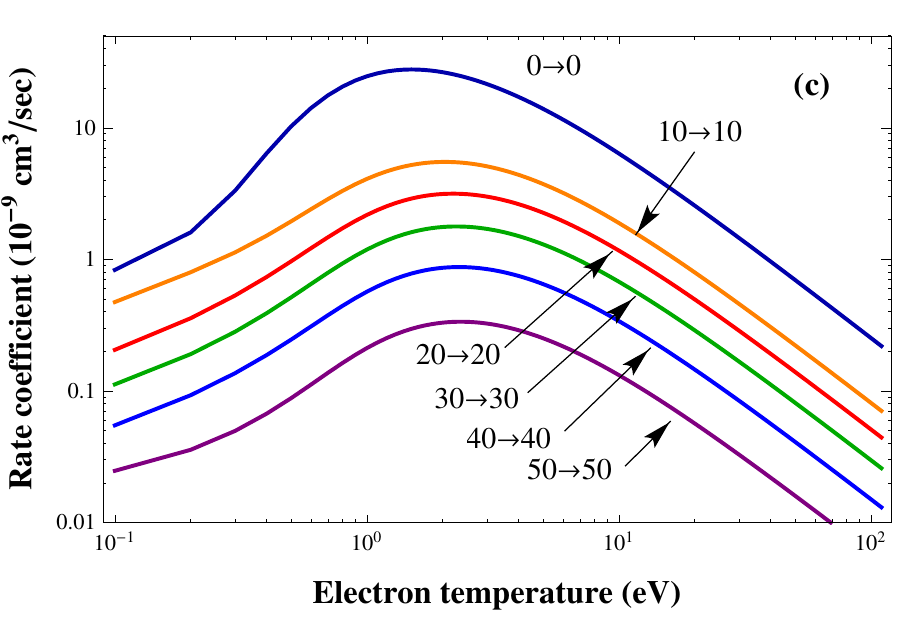}\hfill
\includegraphics[scale=.7,angle=0]{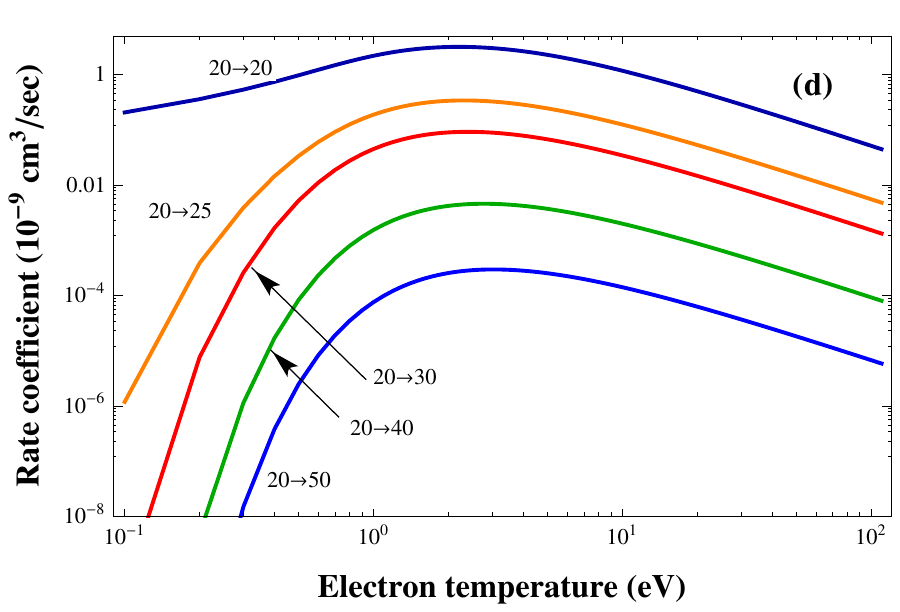}
\end{indented}
\caption{N$_2$ vibrational excitation cross sections and corresponding rate coefficients for (a, c) $v_i=v_f$ and (b, d) $v_i=20 \rightarrow v_f\geq v_i$. }
\label{fig: N2vi-vf}
\end{figure}

RVE cross sections for the inelastic rotationless transitions $v_i=0 \rightarrow v_f=1-5,10$ for N$_2$ are shown in Figs.~\ref{fig: Allan}-\ref{fig: Rmatrix}. A general reduction of the magnitude of the main peaks is observed as the final level is increased. The same trend is exhibited by the curves of Fig.~\ref{fig: N2vi-vf} (a)-(b) where the elastic and $v_i=20\rightarrow v_f\geq v_i$ cross sections, for selected levels $v_f$ of N$_2$, are shown. The trend of the cross sections is reflected in the behavior of the rate coefficients, shown in Figs.~\ref{fig: N2vi-vf} (c)-(d) as a function of the electron temperature, for the same excitation processes as those shown in panels (a)-(b).

\begin{figure}
\begin{indented}
\item[]
\includegraphics[scale=.7,angle=0]{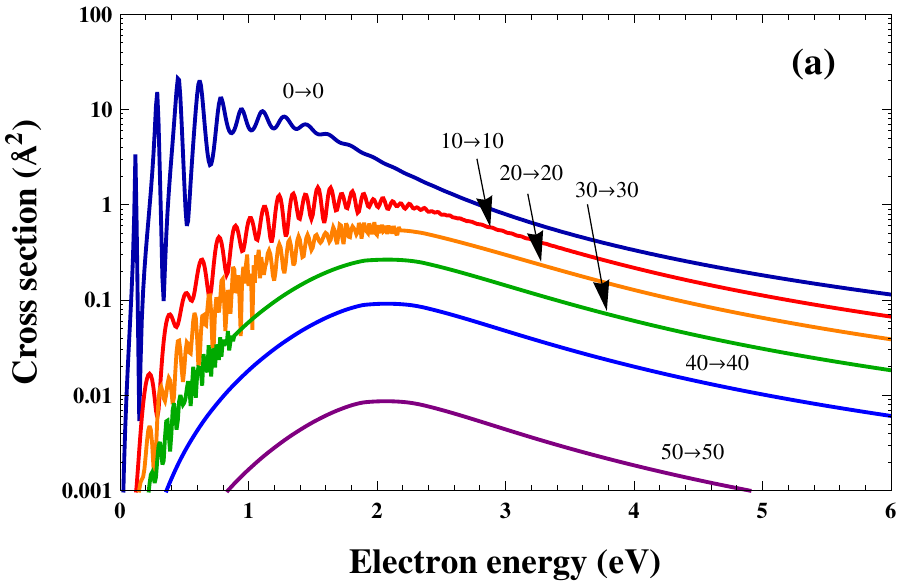}\hfill
\includegraphics[scale=.7,angle=0]{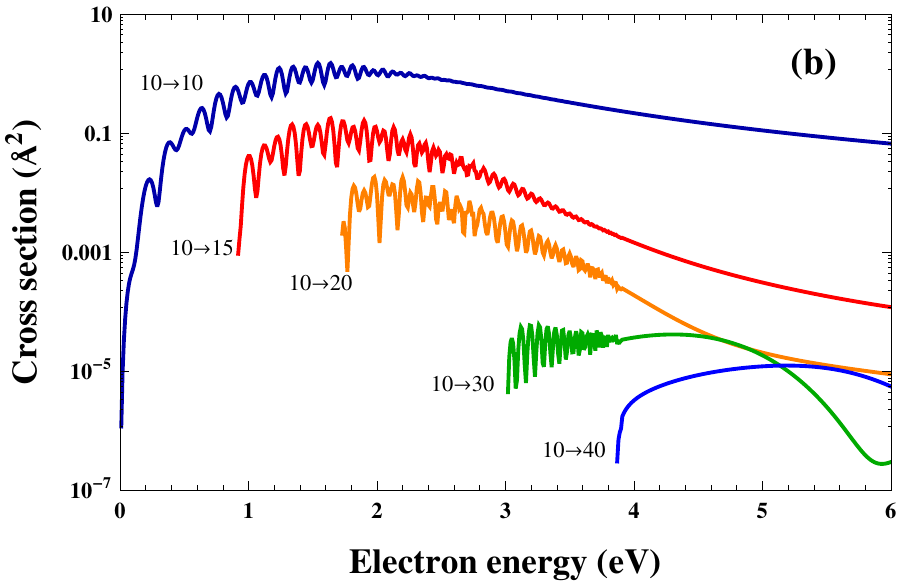}
\\
\includegraphics[scale=.7,angle=0]{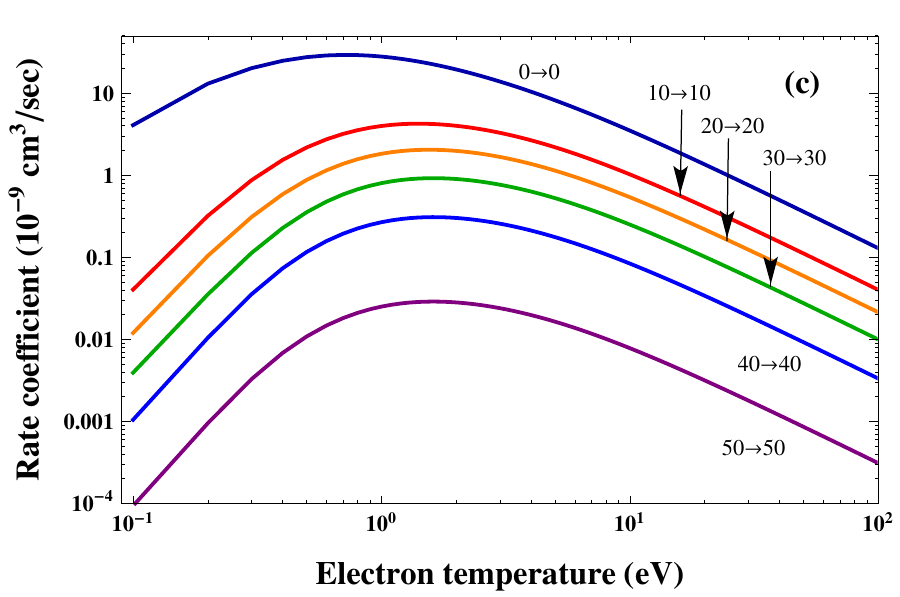}\hfill
\includegraphics[scale=.7,angle=0]{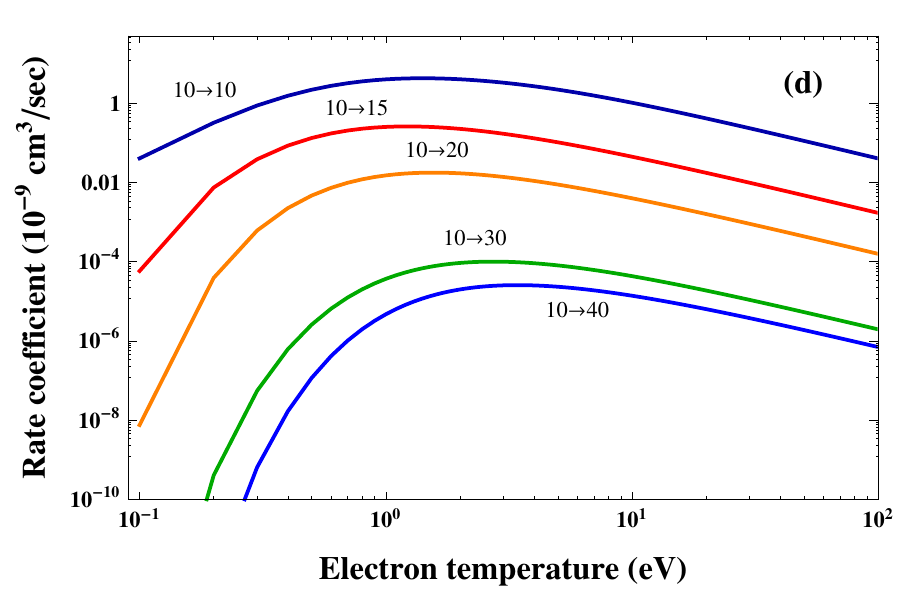}
\end{indented}
\caption{NO excitation cross sections and corresponding rate coefficients for (a, c) $v_i=v_f$ and (b, d) $v_i=10 \rightarrow v_f\geq v_i$.}
\label{fig: NOvi-vf}
\end{figure}

Analogous comments hold for Figs.~\ref{fig: NOvi-vf} where the elastic and inelastic cross sections as well as rate coefficients for process (\ref{RVE NO}), involving the NO molecule, are shown for the rotationless vibrational transitions indicated in the figures.

\begin{figure}
\begin{indented}
\item[]
\includegraphics[scale=.7,angle=0]{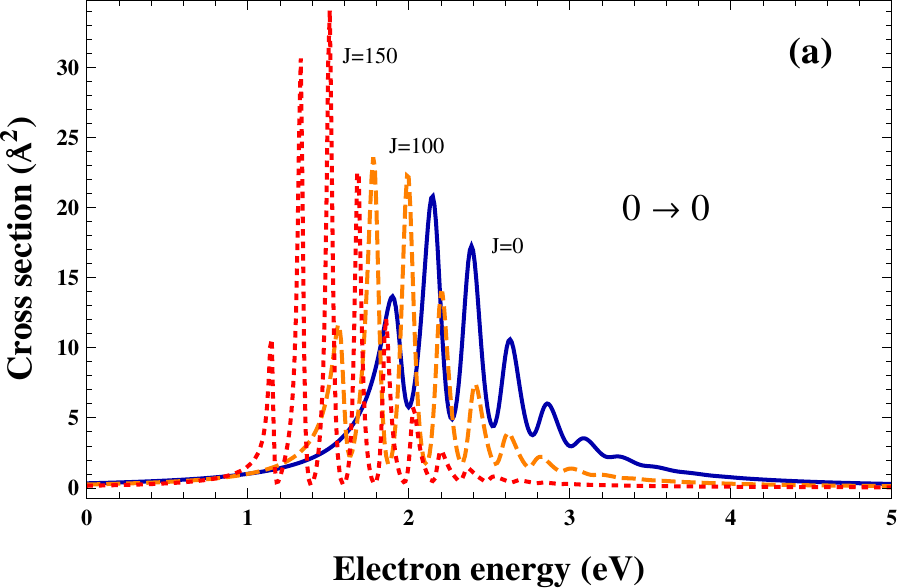}\hfill
\includegraphics[scale=.7,angle=0]{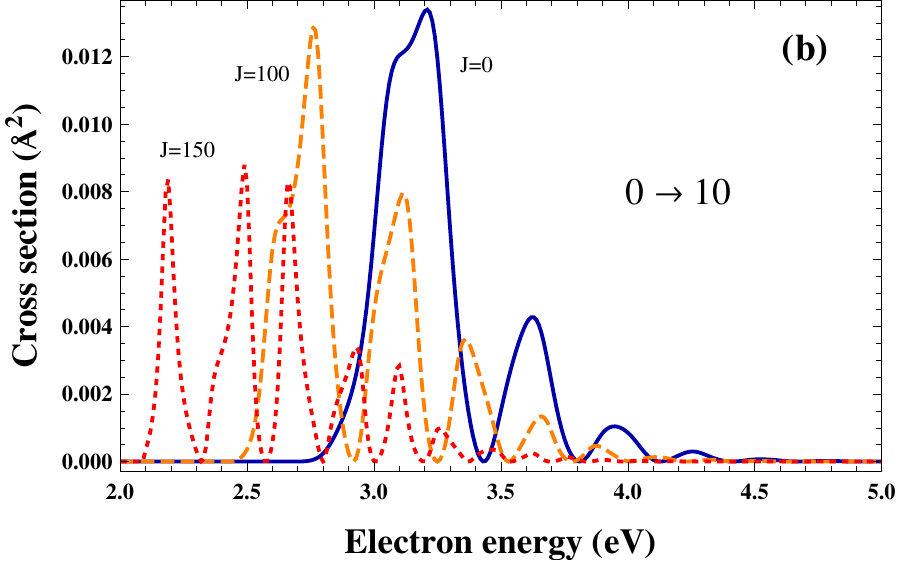}
\\
\includegraphics[scale=.7,angle=0]{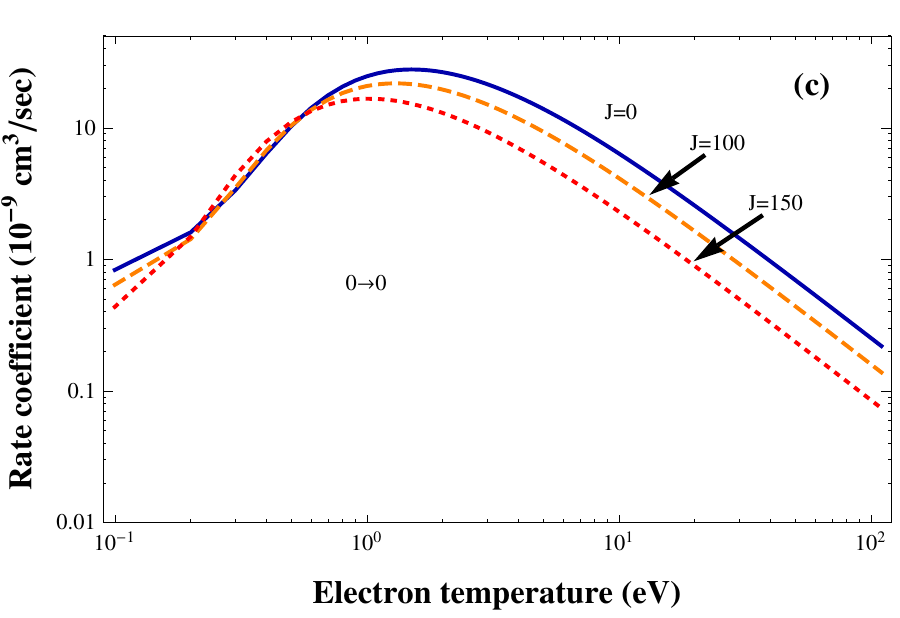}\hfill
\includegraphics[scale=.7,angle=0]{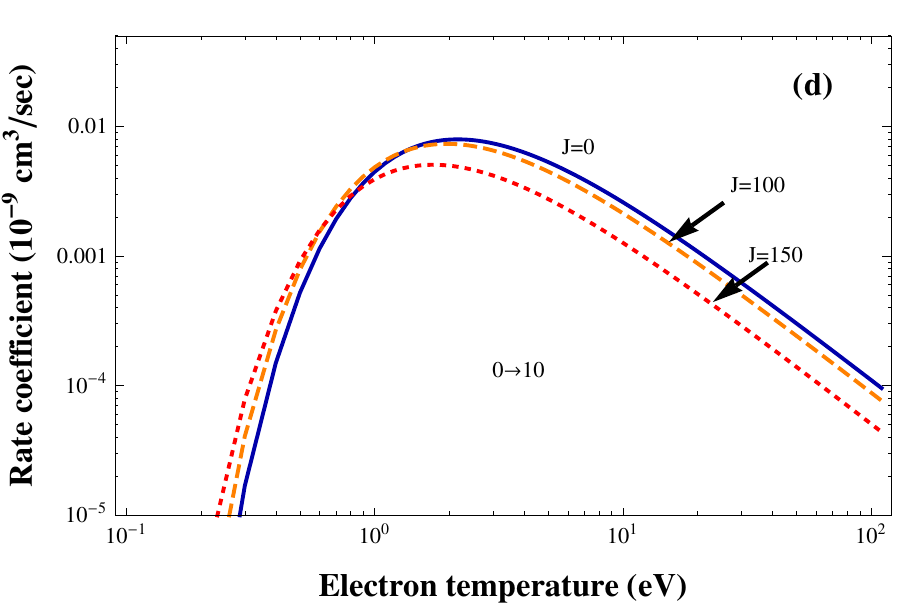}
\end{indented}
\caption{Cross sections and corresponding rate coefficients for (a, c) $0 \rightarrow 0$ elastic  and (b, d) $0\rightarrow 10$ inelastic  vibrational transitions involving the N$_2$ molecule initially in different rotational levels: full line $J$ = 0, dashed line $J$ = 100, dotted line $J$ = 150.}
\label{fig: N2J}
\end{figure}

\begin{figure}
\begin{indented}
\item[]
\includegraphics[scale=.7,angle=0]{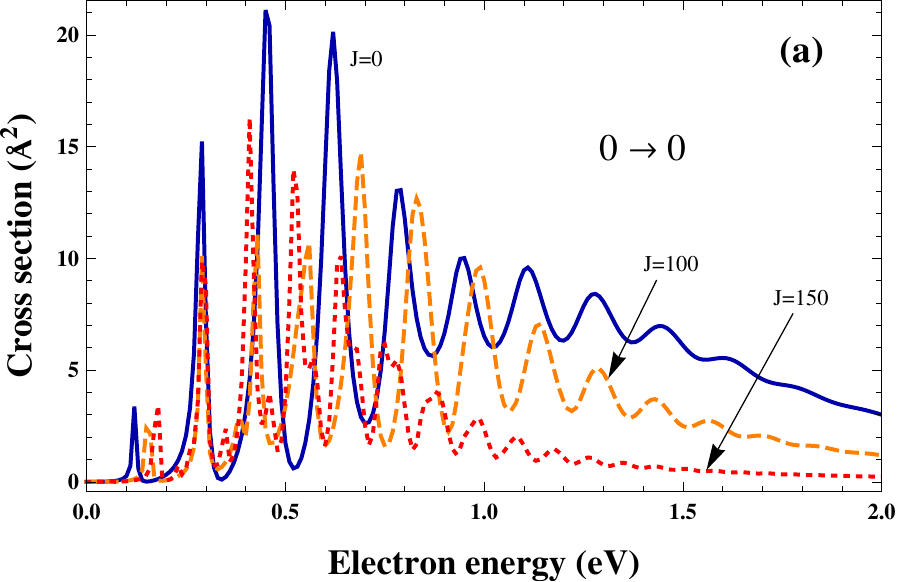}\hfill
\includegraphics[scale=.7,angle=0]{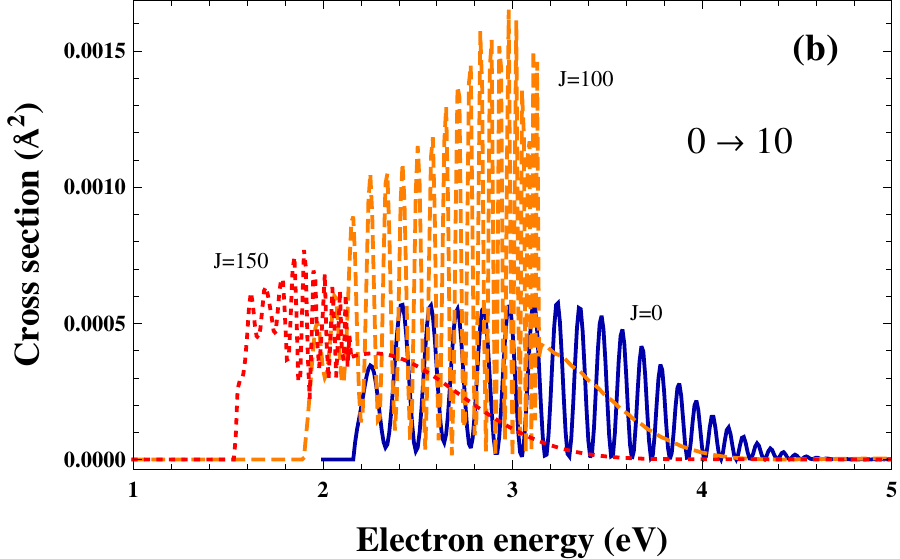}
\\
\includegraphics[scale=.7,angle=0]{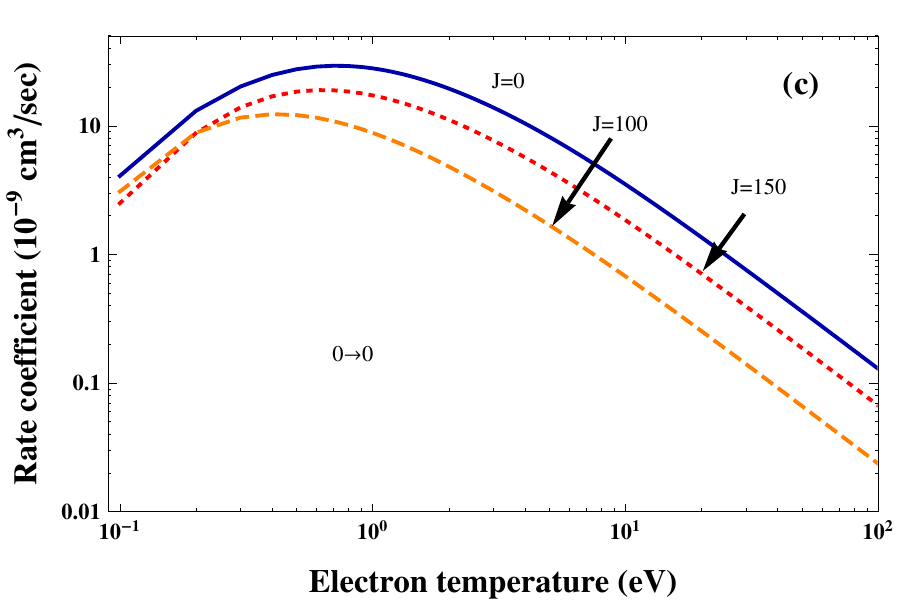}\hfill
\includegraphics[scale=.7,angle=0]{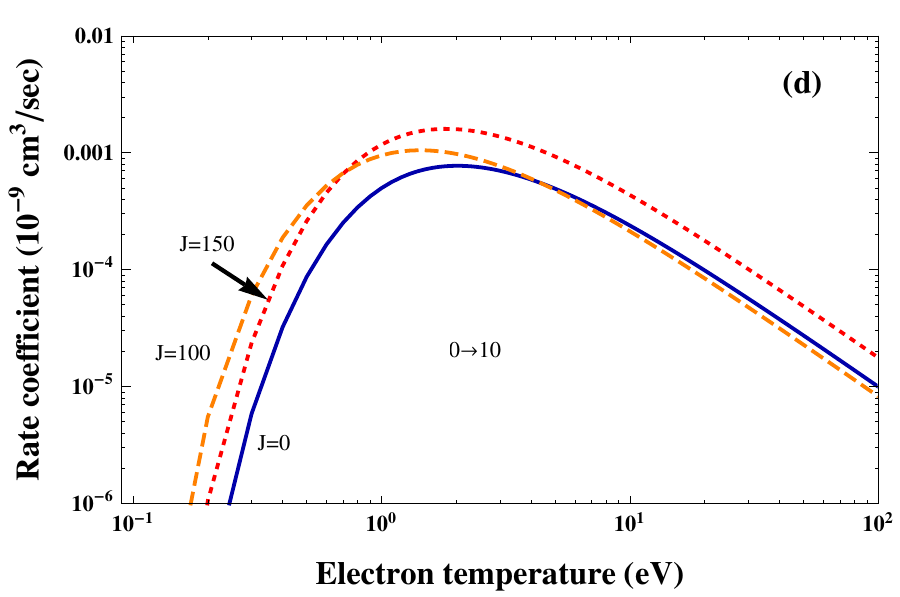}
\end{indented}
\caption{Same as Fig.~\ref{fig: N2J} for e-NO process.}
\label{fig: NOJ}
\end{figure}

In view of the plasma modeling applications, we have also investigated the effect of the rotational excitation of molecule on cross sections and rates. Fig.~\ref{fig: N2J} (a)-(b) shows the cross sections for the two vibrational transitions $0\rightarrow0$ and $0\rightarrow10$, respectively, occurring through process (\ref{RVE N2}) where now the nitrogen molecule is assumed to be initially as well as finally in different rotational levels ($J_i=J_f=J$). We have in particular considered the three cases where $J=$ 0, 100 and 150. No significant variation is observed in the order of magnitude of the cross sections in going from $J=$ 0 to $J=$ 150. An enhancement of the main peaks, by about a factor of 2, is seen in fact for the vibrationally elastic process in Fig.~\ref{fig: N2J} (a), while a decreasing of the peak values occurs for the vibrationally inelastic transition in Fig.~\ref{fig: N2J} (b) as $J$ is increased. A shift toward smaller energy is also observed, as expected, in both the excitations, coming from the lowering of the threshold caused by a reduction of the transition energy. Fig.~\ref{fig: NOJ} (a)-(b) shows the same cross sections occurring through process (\ref{RVE NO}) for NO molecule. However, in this case the main peaks are reduced for vibrationally elastic process and enhanced for vibrationally inelastic process, with respect to $J=0$, as the value of $J$ is increased.

In the Figs.~\ref{fig: N2J} (c)-(d) and~\ref{fig: NOJ} (c)-(d) are shown the rate coefficients for N$_2$ and NO, respectively, for different $J$ values and for the elastic $0\rightarrow 0$ and inelastic $0\rightarrow 10$ vibrational transitions. Inspection of the figures for the elastic case shows that the rates are more sensitive to the rotational contribution for temperatures above 1 eV for nitrogen and above 0.2 eV for nitric oxide. For $T_e< 1$ eV practically no rotational effects is observed in N$_2$, while for NO the rates become insensitive to the initial rotational excitation of the molecule for high $J$ and for $T_e< 0.2$ eV. For the inelastic case the situation is less clear, but no large variations, as $J$ changes, are observed in the whole range of temperatures investigated.

\section{Summary\label{sec:summary}}

We presented in this paper the vibrational excitation cross sections and rate coefficients for processes occurring through resonant electron-impact on N$_2$ and NO molecules, for modeling purposes. Complete sets of data have been produced by using the local complex potential model for the description of the nuclear dynamics, and by representing the required input electronic potential energies and widths by parametric standard expressions, appropriately adjusted to reproduce the experimental measurements. We also illustrated the calculation by commenting on some particular selected $v_i\rightarrow v_f$ transition, and showing the effect of the initial rotational excitation of the target molecule on RVE cross sections and rates.

The entire cross section and rate coefficient database, including all the $v_i\rightarrow v_f$ transitions, and for $J$ ranging from 0 through 150 is available at the website: http://users.ba.cnr.it/imip/cscpal38/phys4entry/database.html.

\section*{Acknowledgments}

The authors wish to thank Dr. A. Laricchiuta (IMIP-CNR, Bari, Italy), Dr. D. Bruno (IMIP-CNR, Bari, Italy), Dr. T.N. Rescigno (LBNL, Berkeley, USA) for helpful discussions, and Prof. J. Tennyson (UCL, London) for reading the manuscript.

This work has been performed under the auspices of the European Union Phys4Entry project funded under grant FP7-SPACE-2009-1 242311.

\section*{References}

\bibliographystyle{unsrt}{}

\section*{Figures}

\end{document}